\documentclass[twocolumn,twoside]{Jornadas}
\usepackage[utf8]{inputenc}
\usepackage[spanish]{babel}
\usepackage{listings}
\usepackage{algorithm}
\usepackage{algorithmicx}
\usepackage{algcompatible}
\usepackage{adjustbox}
\usepackage{graphicx}
\usepackage{color}
\usepackage{caption}
\usepackage[caption=false,font=footnotesize]{subfig}
\usepackage{placeins}
\usepackage{hyperref}
\usepackage{url}
\usepackage{microtype}

\definecolor{gray97}{gray}{.97}
\definecolor{gray75}{gray}{.75}
\definecolor{gray45}{gray}{.45}

\def\BibTeX{{\rm B\kern-.05em{\sc i\kern-.025em b}\kern-.08em
    T\kern-.1667em\lower.7ex\hbox{E}\kern-.125emX}}

\graphicspath{{.}{./figures/}}

\begin{document}

\title{RRCD: Redirección de Registros Basada en Compresión de Datos para Tolerar Fallos Permanentes en una GPU}



\author{%
     Yamilka Toca-Díaz%
     \thanks{División de Aplicaciones y Servicios Informáticos de Geocuba, Universidad de Camagüey, Cuba, e-mail: {\tt yamilka@geocuba.cu.}},
     Alejandro Valero$^2$, Rubén Gran-Tejero$^2$ y Darío Suárez-Gracia%
     \thanks{Dpto. de Informática e Ingeniería de Sistemas, Universidad de Zaragoza, España, 
     e-mail: {\tt \{alvabre,rgran,dario\}@unizar.es.}}
}

\maketitle
\markboth{}{}
\pagestyle{empty} 
\thispagestyle{empty} 

\begin{abstract}
La creciente demanda de paralelismo de las aplicaciones de propósito general en unidades de procesamiento gráfico (GPU) empuja hacia
bancos de registros cada vez más grandes y con un mayor consumo energético en sucesivas generaciones. Reducir la tensión de alimentación más allá de su límite de seguridad es una forma eficaz de mejorar la eficiencia energética del banco de registros. Sin embargo, operar en estas tensiones tan bajas compromete la fiabilidad del circuito. Este trabajo tiene como objetivo tolerar fallos permanentes debidos a variaciones en el proceso de fabricación del banco de registros de una GPU operando por debajo del límite de seguridad. Para ello, este artículo propone una técnica microarquitectónica de redirección de registros, RRCD, que aprovecha la redundancia de los datos inherente en las aplicaciones para comprimir registros en tiempo de ejecución y sin la asistencia del compilador ni modificaciones en el repertorio de instrucciones. En lugar de deshabilitar toda una entrada de registro defectuosa, RRCD aprovecha las celdas fiables en una entrada defectuosa para redirigir y almacenar registros comprimidos. Los resultados experimentales muestran que, con más de un tercio de entradas de registro defectuosas, RRCD asegura la fiabilidad del banco de registros y reduce el consumo de energía en un 47\% con respecto a un diseño convencional alimentado con una tensión nominal. El ahorro de energía es un 21\% en comparación con un esquema de suavizado de ruido de tensión que opera en el límite seguro de tensión. Estos beneficios se obtienen con un impacto en el rendimiento y área del sistema menor que un 2 y 6\%, respectivamente.
\end{abstract}


\maketitle

\section{Introducción}

\PARstart{D}{esde} hace más de una década, las unidades de procesamiento gráfico (GPUs) se han consolidado como dispositivos de computación masivamente paralelos y se han adoptado en múltiples áreas de la computación, desde sistemas empotrados hasta grandes centros de datos de alto rendimiento. Este éxito ha resultado en una gran cantidad de aplicaciones de propósito general programadas y optimizadas para su ejecución en una GPU (aplicaciones GPGPU). Entre ellas, las aplicaciones emergentes, como las relativas al aprendizaje profundo, la analítica o la minería de datos, empujan hacia el diseño de arquitecturas GPU con mayores recursos computacionales y de almacenamiento mientras se mantiene el consumo de energía bajo control. En consecuencia, la eficiencia energética se ha convertido en uno de los principales aspectos de diseño de la arquitectura de GPUs modernas.

El banco de registros es una de las estructuras de memoria que más energía consume en una GPU, siendo responsable de aproximadamente un 20\% del consumo total de energía del dispositivo~\cite{access} y su consumo aumenta generación tras generación. Por ejemplo, el banco de registros de NVIDIA Tesla V100, con 20 MB, es 5 veces más grande que su homólogo en Tesla K40~\cite{Volta2017}. Muchas propuestas se han centrado en la eficiencia energética del banco de registros, desde técnicas tradicionales, como \emph{clock} y \emph{power gating}, hasta propuestas recientes, incluyendo la explotación de patrones de acceso a los datos~\cite{energy2}, el análisis del tiempo de vida de los datos~\cite{energy}, técnicas de prebúsqueda~\cite{ltrf}, redundancia de los datos~\cite{wctc}, uso compartido de registros~\cite{virtual} o técnicas de \emph{coalescing}~\cite{corf}. 

Por otro lado, como cualquier otro circuito digital, el banco de registros de la GPU se encuentra afectado por los efectos de variación estática y dinámica. Las variaciones estáticas son una consecuencia del proceso de fabricación del chip, mientras que las variaciones dinámicas provienen del funcionamiento del circuito (e.g., ruido de tensión y efectos de envejecimiento). Las variaciones en el proceso de fabricación imponen un límite mínimo de tensión de alimentación de seguridad ($V_{min}$) a cada celda de memoria para garantizar su fiabilidad y, a su vez, para un banco de registros completo, $V_{min}$ queda establecida como la $V_{min}$ más alta correspondiente a la peor celda.

Un suministro de tensión ($V_{dd}$) por encima de $V_{min}$ asegura un margen de protección suficiente para una operación más segura ante de una caída repentina de $V_{dd}$, pero por el contrario, acelera el envejecimiento del circuito. Además, una $V_{dd}$ alta supone un desperdicio de energía, puesto que la energía escala cuadráticamente con $V_{dd}$ y un ruido notable en la tensión de alimentación es un evento poco frecuente~\cite{dvfs_gpu2}.
En este contexto, algunos trabajos anteriores han propuesto esquemas de suavizado del ruido de tensión para el banco de registros de la GPU, los cuales permiten relajar el margen de protección empujando $V_{dd}$ hacia $V_{min}$ con una frecuencia fija~\cite{dvfs_gpu,dvfs_gpu2}. Reducir $V_{dd}$ por debajo de $V_{min}$ es una tarea desafiante debido al elevado número de fallos permanentes que pueden producirse como resultado de superar la $V_{min}$ de múltiples celdas~\cite{salami}. A diferencia de los fallos solitarios e inducidos por las variaciones dinámicas o el impacto de partículas, el elevado número de fallos permanentes que surgen cuando $V_{dd}< V_{min}$ está lejos de las capacidades de los códigos de corrección de errores (ECC) convencionales, lo cual no sólo requiere de mayores capacidades de almacenamiento y consumo energético sino también de (de)codificadores lentos y complejos para garantizar una operación segura sobre el banco~\cite{ecc2,ecc}.

Con el objetivo de mejorar aún más la eficiencia energética sin utilizar códigos ECC costosos, se está explorando la redirección de contenidos como una solución para tolerar fallos permanentes debido a las variaciones en el proceso de fabricación del banco de registros GPU operando en una tensión por debajo de $V_{min}$. Este enfoque consiste en deshabilitar las entradas de registro defectuosas y proporcionar entradas alternativas y fiables hacia donde los accesos a registros defectuosos se redirigen. En este contexto, el trabajo previo GR-Guard identifica entradas fiables que contienen datos inútiles en tiempo de compilación y redirige accesos defectuosos a dichas entradas en tiempo de ejecución gracias a modificaciones en el repertorio de instrucciones (ISA)~\cite{Tan2016}. Sin embargo, modificar la ISA presenta los siguientes inconvenientes: i) expone detalles de implementación innecesarios al programador, ii) requiere cambios en el software para explotar el mecanismo y iii) aumenta la complejidad de la ISA (compatibilidad con versiones anteriores y extensiones futuras). Por el contrario, este trabajo presenta una técnica novedosa de redirección a nivel de microarquitectura, RRCD, que habilita entradas defectuosas con fines de redirección al explotar la compresión de datos inherente de las aplicaciones GPGPU en tiempo de ejecución.

Para que la compresión de datos sea efectiva, se requiere la presencia de patrones regulares en el flujo de datos. Afortunadamente, muchos lenguajes de programación para GPU generan patrones de acceso a memoria regulares y evitan la divergencia en el flujo de ejecución, almacenando patrones de datos regulares en los registros de la GPU~\cite{xiang2013}. Estos patrones regulares se pueden comprimir utilizando variaciones del algoritmo \emph{Base-Delta-Immediate} (BDI) originalmente propuesto para comprimir líneas de cache de la CPU~\cite{bdi}. Estos algoritmos de compresión se pueden aplicar a cualquier arquitectura GPU de NVIDIA o AMD, ya que los patrones de datos explotados provienen exclusivamente de modelos de programación de una instrucción-múltiples hilos proporcionado por CUDA u OpenCL.

Algunos trabajos recientes explotan la compresión de datos en el banco de registros de la GPU para reducir energía~\cite{wctc,ereer,sttcomp}, mitigar efectos de envejecimiento de los transistores~\cite{valerotc} o para lidiar con fallos transitorios~\cite{mittal}. Sin embargo, a nuestro mejor entender, la compresión no se ha utilizado con anterioridad para eludir fallos permanentes en bancos de registros GPU. Además, este es el primer trabajo que propone un mecanismo de redirección sin introducir complejidad o una sobrecarga en el software o el programador.


Los resultados experimentales muestran que RRCD garantiza un funcionamiento fiable de un banco de registros con un 39\% de entradas de registro defectuosas, reduciendo el consumo de energía promedio en un 47 y un 21\% con respecto a un banco de registros convencional que opera con una $V_{dd}$ nominal y con una $V_{min}$ segura, respectivamente, mientras que el impacto en el rendimiento y área es inferior al 2 y 6\%, respectivamente.


\section{Antecedentes}
\label{bckgnd}

Esta sección resume la arquitectura del banco de registros de la GPU, el modelo de fiabilidad y los escenarios utilizados para evaluar la propuesta, así como la estrategia de compresión de datos utilizada para combatir las variaciones en el proceso de fabricación.

\subsection{Banco de Registros GPU}

Las GPUs actuales constan de decenas de procesadores en orden, también conocidos como \emph{Streaming Multiprocessors} (SMs) y \emph{Compute Units} (CUs) en las GPUs de NVIDIA y AMD, respectivamente. Sin pérdida de generalidad, este trabajo utiliza como ejemplo la familia de GPUs AMD Graphics Core Next (GCN)~\cite{amdwhite}. Por tanto, el resto del documento utilizará la terminología de AMD.

\begin{figure}[t!]
\centering
\includegraphics[width=0.99\columnwidth]{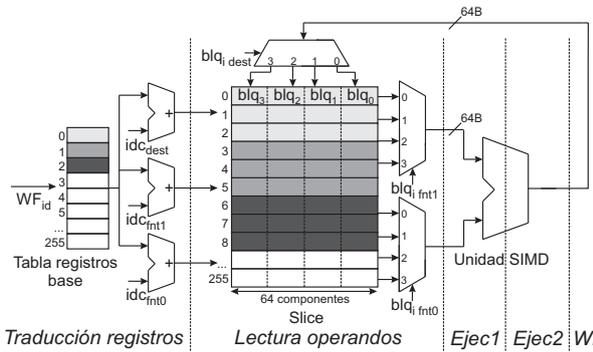}
\caption{Pipeline correspondiente a una unidad SIMD después de las etapas de búsqueda y decodificación.}
\label{pipe}
\end{figure}

Una CU consta de 4 unidades \emph{single-instruction, multiple-data} (SIMD). Cada unidad SIMD se asocia con un segmento o \emph{slice} de 64 KB del banco de registros. La Figura~\ref{pipe} muestra las etapas de \emph{pipeline} asociadas a una unidad SIMD y cómo 256 entradas de registros vectoriales componen un slice. A su vez, cada entrada consta de 64 componentes de 4 bytes. Para acceder a las entradas de registro, los hilos se organizan en grupos de hasta 64 hilos llamados \emph{wavefronts}. Todos los hilos que pertenecen a un mismo \emph{wavefront} acceden a la misma entrada de registro, pero con un desplazamiento de componente basado en el identificador de hilo en el \emph{wavefront}. De esta forma, aunque cada hilo trabaja con una componente diferente de la misma entrada, se evita hacer referencia a cada componente en la ISA.

Dado que una unidad SIMD consta de 16 canales (64 B), los hilos de un \emph{wavefront} ejecutan una instrucción formando 4 paquetes de 16 hilos adyacentes, denominados como \emph{bloques} ($blq_i$) en la figura. Estos bloques se acceden uno tras otro durante cuatro ciclos sucesivos. Dado que las instrucciones de AMD suelen disponer de dos operandos fuente ($fnt0$ y $fnt1$) y un operando de destino ($dest$), un slice incorpora dos puertos de lectura y un puerto de escritura.

Las entradas de registro de un slice se distribuyen estáticamente entre los \emph{wavefronts} que se ejecutan en la unidad SIMD correspondiente. Para ello, cuando un \emph{wavefront} se asigna a una unidad SIMD, se le añade un identificador ($WF_{id}$), la dirección física de su registro base y un número de entradas de registro contiguas, siendo este último un valor constante para todos los \emph{wavefronts} de una aplicación~\cite{amd2015}. El conjunto de entradas de un \emph{wavefront} se denomina ventana de registros. Por ejemplo, la Figura~\ref{pipe} muestra ventanas de 3 registros en el slice, resaltadas con un gris de mayor gradiente para los \emph{wavefronts} $WF_{0}$, $WF_{1}$ y $WF_{2}$.

Como se muestra en la etapa de traducción de registros, las instrucciones de cada \emph{wavefront} se refieren a registros lógicos dentro de su ventana con índices ($idc$). Estos índices se agregan al registro base para obtener las entradas de registro físico. La entrada del registro base se obtiene de una tabla base de registros indexada con el id del \emph{wavefront}.

\subsection{Escenarios de Fiabilidad}
\label{rel}



La distribución de fallos permanentes en una memoria SRAM depende del porcentaje de impacto de las variaciones sistemáticas (transistores vecinos con variaciones similares) y aleatorias (variaciones uniformes a lo largo del chip). La propuesta se evalúa bajo tres escenarios de fiabilidad: \emph{Común}, \emph{Agrupado} y \emph{Disperso}. El escenario \emph{Común} se refiere a un escenario ampliamente utilizado donde las variaciones sistemáticas y aleatorias se tratan por igual~\cite{variusnvt}, mientras que \emph{Agrupado} y \emph{Disperso} hacen referencia a escenarios con un mayor impacto de variaciones sistemáticas y aleatorias, respectivamente.

\begin{table}[t!]
\renewcommand{\baselinestretch}{1}
\small
\centering
\caption{Porcentaje de entradas de registro con diferente número de bits defectuosos para cada escenario de fiabilidad~\protect\cite{Tan2016}.}
{
\begin{tabular}{|c||c|c|c|c|c|} \hline
Escenario &  \multicolumn{2}{c|}{Entradas fiables} & \multicolumn{3}{c|}{Entradas defectuosas}\\
de fiabilidad     &  0-bit & 1-bit & 2-bit & 3-bit & $\geq$4-bit\\\hline\hline
\emph{Común}  & 34 & 33 & 20 & 10 & 3   \\\hline
\emph{Agrupado} & 43 & 20 & 12 & 10 & 15  \\\hline
\emph{Disperso} & 26 & 35 & 23 & 12 & 4  \\\hline
\end{tabular}}
\label{fop}
\end{table}

La Tabla~\ref{fop} muestra la distribución de las entradas de registro de un slice según su número de bits defectuosos para cada escenario de fiabilidad. Este trabajo asume el modelo de fallos de 28 nm propuesto en~\cite{Tan2016} y obtenido con VARIUS~\cite{iccd16}.

El modelo de fiabilidad asumido se centra en el banco de registros, asumiendo una implementación con dominios de tensión dedicados para lógica y memoria, manteniendo la lógica con una $V_{dd}$ elevada para evitar fallos como se describe en~\cite{vdomains,vdomains2}.
$V_{dd}$ queda establecida durante toda la ejecución de una aplicación en 419, 497 y 371 mV para los escenarios \emph{Común}, \emph{Agrupado} y \emph{Disperso}, respectivamente, todas ellas por debajo de una $V_{min}$ de 600 mV~\cite{vmin,Tan2016,Ipatch}. En aras de mejorar la claridad, se han agrupado las entradas con cuatro o más bits defectuosos.

\emph{Error-Correcting Pointer} (ECP) es un enfoque que corrige fallos permanentes mediante la codificación de las ubicaciones de los bits defectuosos en una tabla y la asignación de bits de reemplazo adicionales para reemplazar a los bits defectuosos~\cite{ecp}. En este trabajo, ECP se emplea con una granularidad razonable de entrada de registro con un bit de reemplazo por entrada. Por tanto, las entradas con menos de dos bits defectuosos se consideran como fiables. De manera conservadora, asumimos que los bits defectuosos se distribuyen uniformemente entre los cuatro bloques de una entrada de registro. Es decir, las entradas defectuosas con $i$ bits defectuosos, $i\geq2$, tienen $i$ bloques defectuosos. Por supuesto, cuando $i\geq4$, todos los bloques son defectuosos y la entrada se considera como completamente inútil.

 El escenario \emph{Común} muestra una distribución de registros donde el porcentaje de entradas se reduce con el número de bits defectuosos por entrada. En contraste, el escenario \emph{Agrupado}, donde el efecto sistemático domina sobre el efecto aleatorio, muestra un mayor porcentaje de entradas en los extremos de cero y cuatro o más bits defectuosos. El escenario \emph{Disperso}, donde los fallos se distribuyen aleatoriamente, presenta un mayor porcentaje de entradas con al menos un bit defectuoso, pero no muestra tantas entradas completamente defectuosas como el escenario \emph{Agrupado}. En general, el porcentaje de entradas defectuosas en un slice es 33, 37 y 39\% para los escenarios \emph{Común}, \emph{Agrupado} y \emph{Disperso}, respectivamente.

Finalmente, cada escenario de fiabilidad cuenta con mapas de fallos de $256\times4$ bits por slice de acuerdo con la ubicación de los bloques defectuosos. Estos mapas se determinan durante una prueba posterior a la fabricación~\cite{Ipatch,Tan2016,tanmap} y se utilizarán como una entrada de RRCD para diferenciar entre bloques (y entradas) de registro fiables y defectuosos.

\subsection{Compresión de Datos}
\label{datac}

Este artículo explota el mecanismo de compresión de datos propuesto en~\cite{valerotc}, donde se identifican una serie de patrones de datos regulares en el banco de registros GPU. Estos patrones aparecen cuando todas las componentes de un registro almacenan el mismo valor escalar debido al control de divergencias~\cite{xiang2013}, cuando un registro almacena una secuencia de valores donde la diferencia entre componentes consecutivas es constante debido a identificadores de hilo o direcciones de vectores y cuando, aparte de esta diferencia mencionada, aparece una segunda diferencia de valor entre otras componentes adyacentes, siendo el caso de registros que almacenan linealmente las direcciones de una matriz. Este último patrón también aparece cuando se utilizan técnicas de programación como \emph{tiling} o ventanas deslizantes.

Estos patrones ofrecen una relación de compresión muy alta, puesto que se necesitan tan sólo 4,88 B para (des)comprimir una entrada de registro de 256 B. Las unidades de hardware requeridas para (des)comprimir un registro en tiempo de ejecución consisten en una serie de sumadores, restadores, comparadores y pequeños \emph{buffers} de memoria para hacer frente a las operaciones de lectura y escritura de cuatro ciclos en un slice de acuerdo con el número de bloques en una entrada de registro. En este sentido, una unidad de descompresión recibe un bloque con los datos comprimidos de un registro y envía cada bloque sin comprimir en un ciclo sucesivo, completando el proceso tras cuatro ciclos. De manera similar, una unidad de compresión recibe cada bloque sin comprimir uno tras otro, obteniendo la potencial compresión en el primer ciclo y determinando si el registro completo se puede comprimir en el cuarto ciclo tras examinarse todos los bloques.

\section{Motivación}
\label{mot}

Esta sección explora el potencial de una técnica de redirección de registros basada en la compresión de datos para tolerar fallos permanentes en el banco de registros GPU. Para ello, las entradas de registros se clasifican dependiendo del estado de la compresión de los contenidos (compresible o incompresible) y el estado de la fiabilidad de la entrada (fiable o defectuosa).

\begin{figure}[t!]
\centering
\includegraphics[width=0.99\columnwidth]{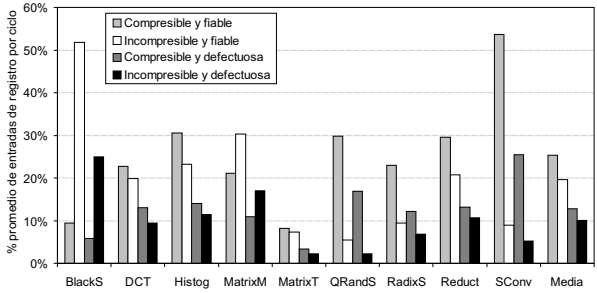}
\caption{Porcentaje promedio de entradas de registro por ciclo según su estado de fiabilidad y compresión de contenidos.}
\label{motfig}
\end{figure}

La Figura~\ref{motfig} muestra las cuatro categorías resultantes. Cada barra muestra el porcentaje promedio de entradas ocupadas por ciclo. Los resultados se limitan al análisis del escenario de fiabilidad \emph{Común} y se evalúa un subconjunto representativo de aplicaciones de la \emph{suite} OpenCL SDK 2.5 con características diferentes como requisitos de memoria, estrés del banco de registros y oportunidades de compresión de datos~\cite{opencl}. Refiérase a la Sección~\ref{exp} para más detalles sobre el entorno experimental.

Casi un 30\% de las entradas son fiables y almacenan un registro comprimido por defecto en cada ciclo. Estas entradas son candidatas  principales para la redirección, puesto que la técnica RRCD trata de forzar la redirección de un registro incompresible hacia una entrada fiable. Los resultados también muestran que un 22\% de las entradas son fiables y almacenan un registro sin comprimir. Por tanto, no constituyen entradas apropiadas para la redirección. Un porcentaje reducido del 10\% de las entradas son defectuosas y almacenan un registro sin comprimir. Un registro en una entrada defectuosa requiere una redirección hacia una entrada fiable, de lo contrario, la integridad de los datos estaría comprometida con la presencia de fallos. Finalmente, un 14\% de las entradas son defectuosas pero almacenan un registro comprimido. En este caso, no se necesitaría una redirección siempre y cuando el registro comprimido esté almacenado en un bloque fiable dentro de la entrada defectuosa. Los bloques fiables restantes (si los hubiese) de dicha entrada pueden almacenar otros registros comprimidos, lo que da como resultado una entrada defectuosa que almacena múltiples registros comprimidos.

En general, la compresión de datos ofrece oportunidades prometedoras para la técnica de redirección, ya que todas las aplicaciones, aparte de \emph{BlackS}, muestran un mayor porcentaje de entradas fiables que almacenan registros comprimidos frente a entradas defectuosas que almacenan registros sin comprimir. Nótese que las entradas no asignadas también ofrecen oportunidades de redirección para registros comprimidos o sin comprimir. La utilización del banco de registros depende de: i) la arquitectura de la GPU, incluyendo el número de entradas del slice, el número máximo de \emph{wavefronts} concurrentes asignados a un slice y el tamaño de la ventana de registros~\cite{amd2015} y ii) optimizaciones de las aplicaciones~\cite{abdel}. En la Figura~\ref{motfig}, la suma de cada categoría para una aplicación determinada constituye el porcentaje de utilización del banco de registros. Dependiendo de la aplicación, estos porcentajes oscilan entre 54 (\emph{QRandS}) y 93\% (\emph{SConv}), con un promedio del 74\%, siendo un porcentaje más alto que en estudios anteriores~\cite{abdel}.

\begin{figure*}[t!]
\centering
\includegraphics[width=0.99\textwidth]{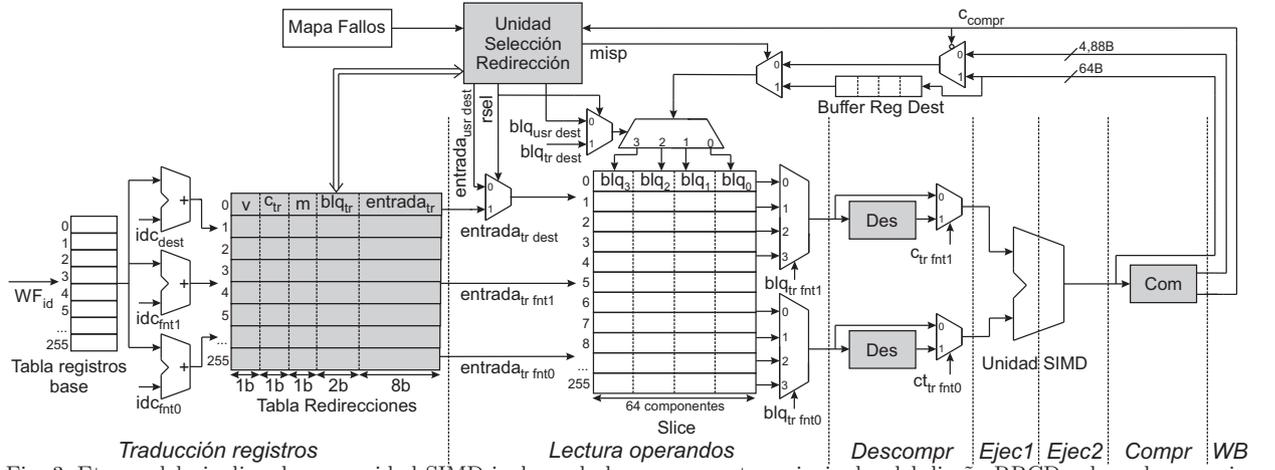}
\caption{Etapas del pipeline de una unidad SIMD incluyendo los componentes principales del diseño RRCD coloreados en gris.}
\label{dcpatch}
\end{figure*}

\section{Propuesta RRCD}
\label{prop}
Esta sección presenta la propuesta RRCD. En primer lugar, se presenta una visión general de RRCD, discutiendo cómo funciona el algoritmo de selección de redirecciones e identificando los componentes principales del diseño. A continuación, se describen estos componentes en detalle, incluyendo las operaciones implicadas en cada uno de ellos. Por último, se analizan los costes de temporización, energía, potencia y área.

\subsection{Vista General}

RRCD redirecciona dinámicamente registros con entradas del slice en función del estado de compresión del registro. Se requiere una redirección cuando se escribe un registro por primera vez, así como cuando las escrituras posteriores cambian su estado de compresión. Esto significa que se libera la entrada anterior del registro y se le asigna al mismo una nueva entrada defectuosa o fiable en función de su nuevo estado. Los registros no comprimidos siempre se redireccionan a entradas fiables, mientras que los comprimidos pueden redireccionarse a entradas fiables o defectuosas. En este sentido, RRCD prioriza que los registros comprimidos se asignen a entradas defectuosas. De lo contrario se desperdiciarían entradas fiables.


El enfoque propuesto aprovecha el acceso segmentado de una entrada de registro de 256 bytes en bloques de 64 bytes. Es decir, una entrada fiable puede contener hasta cuatro registros comprimidos, cada uno en un bloque diferente. En cambio, las entradas defectuosas sólo pueden contener registros comprimidos en bloques fiables, mientras que los bloques defectuosos permanecen deshabilitados. De este modo, al contrario que trabajos previos~\cite{Tan2016}, la compresión de datos permite explotar las entradas defectuosas. Nótese que un registro comprimido (4,88 bytes) ocupa una pequeña fracción del bloque de 64 bytes, lo cual deja espacio para redireccionar más de un registro comprimido dentro de un bloque. Sin embargo, esto conllevaría un diseño más complejo con un mayor coste en energía y área.


La Figura~\ref{dcpatch} muestra las etapas del pipeline de una unidad SIMD tras la búsqueda y decodificación de una instrucción, incluyendo los componentes principales del diseño RRCD coloreados en gris. A primera vista, la Tabla de Redirecciones (TR) se encuentra después de la traducción de registros y envía a la siguiente etapa la entrada donde encontrar cada registro.
Las unidades de (des)compresión (Com/Des) tienen etapas dedicadas en el pipeline. Los descompresores se sitúan tras los puertos de lectura del slice y garantizan que la unidad SIMD opere con bloques no comprimidos en cada ciclo. Después de examinar un bloque destino, la unidad de compresión envía los datos comprimidos (si es el caso) al puerto de escritura del slice. Por último, la Unidad de Selección de Redirección (USR) se encuentra en la etapa de \emph{writeback} (WB). Teniendo en cuenta el mapa de fallos del escenario de fiabilidad, la USR emite una nueva redirección cuando cambia el estado de compresión de un registro de destino. A continuación, se describe en detalle cada uno de estos componentes y las operaciones implicadas.


\subsection{Tabla de Redirecciones (TR)}

La TR es una memoria de 416 bytes indexada por los registros físicos fuente y destino de una instrucción. El número de filas de la TR es igual al número de entradas del slice (es decir, 256) y también están dispuestas en ventanas de registros. De hecho, la TR desvincula completamente las ventanas de registros del slice para maximizar la posibilidad de encontrar una redirección. En otras palabras, las redirecciones permiten asignar dinámicamente registros de un \emph{wavefront} en cualquier entrada de slice disponible.


El contenido de una fila de la TR se actualiza por la USR cuando es necesaria una nueva redirección. Cada fila contiene la entrada ($entrada_{tr}$) en la que reside un registro en el slice. Para un registro comprimido, los bits $blq_{tr}$ indican en qué bloque dentro de la entrada se encuentra el registro. Estos bits también se utilizan en los puertos de lectura del slice (bits $blq_{tr \; fnt_{i}}$) para enviar un bloque fuente a la siguiente etapa en un único ciclo. Del mismo modo, cuando el estado de un registro destino permanece como comprimido (es decir, no se requiere una nueva redirección de la USR), los bits $blq_{tr}$ de dicho registro controlan el puerto de escritura (bits $blq_{tr \; dest}$) para almacenar el resultado comprimido de la unidad SIMD en el bloque adecuado del slice en un único ciclo. En el caso de un registro no comprimido, los bits $blq_{tr}$ no se utilizan para controlar los puertos del slice. En su lugar, se accede a todos los bloques del registro en cuatro ciclos consecutivos.


Los bits de validez $v$ y compresión $c_{tr}$ definen el estado de un registro, codificando un registro válido y comprimido con un `1' lógico en los bits. La USR utiliza estos bits para obtener la redirección de un registro destino. Además, el bit $c_{tr}$ acciona los muxes 2:1 de la etapa de descompresión para seleccionar entre bloques no comprimidos de un puerto de lectura o bloques descomprimidos de una unidad Des.


Nótese que si no es posible encontrar una redirección para un registro comprimido o sin comprimir, se asigna una redirección de respaldo al registro (\emph{spill}). Es decir, una estructura de memoria adicional almacena el contenido del registro. En este trabajo, en lugar de contaminar o reducir la capacidad de almacenamiento efectiva de la jerarquía de memoria~\cite{Tan2016}, se explota la cache Local Data Share (LDS)~\cite{amdwhite} como almacenamiento de respaldo\footnote{La LDS se comparte entre las unidades SIMD de una CU. Para aplicaciones GPGPU, el uso de la LDS depende del programador. Las aplicaciones pueden cargar o almacenar datos en la LDS para amplificar el ancho de banda de cache, evitar la contaminación en la jerarquía de memoria con operaciones \emph{scatter} y \emph{gather} o realizar operaciones atómicas a nivel de \emph{work-group}~\cite{amdwhite}.}.


El bit $m$ indica si la redirección de un registro se encuentra en un almacén de respaldo ($m=1$) o en el slice ($m=0$). Si $m=1$, en lugar de acceder al slice, se accede a la cache LDS para obtener los datos solicitados. Para ello, la LDS se divide en dos mitades, una para los datos regulares propios de la LDS y la otra para los registros de respaldo. Estos registos se organizan como un array contiguo en la LDS. En la TR, cuando $m=1$, los bits $entrada_{tr}$ se refieren a un desplazamiento desde la dirección base de la partición de respaldo hasta la dirección donde se encuentra el registro.


\subsection{Unidades de (Des)compresión (Com/Des)}

De acuerdo con el número de puertos del slice, RRCD requiere una unidad Com y dos unidades Des. A diferencia de TR, la latencia de estas unidades obliga a ampliar la profundidad del pipeline en dos etapas adicionales.
La entrada de una unidad Des proviene del puerto de lectura y se refiere a un bloque que contiene los datos comprimidos de un registro fuente. En estos casos, esta unidad desenrolla los cuatro bloques del registro y los envía uno tras otro a la unidad SIMD en ciclos sucesivos. Por el contrario, los bloques procedentes de registros no comprimidos simplemente pasan por alto las unidades Des.


La unidad Com recibe de la unidad SIMD un bloque sin comprimir en cada ciclo. Cuando esta unidad recibe el primer bloque de un registro, determina el estado de compresión del bloque y notifica dicho estado a la USR con el bit $c_{compr}$, con valor `1' o `0' cuando el contenido está comprimido o descomprimido, respectivamente. Si $c_{compr}=1$, el compresor también envía los datos comprimidos hacia el puerto de escritura. En caso contrario, los cuatro bloques de un registro se envían hacia el mismo puerto. El bit $c_{compr}$ también controla un mux 2:1 para enviar los datos comprimidos o no comprimidos hacia el puerto de escritura.


La compresión de un registro es especulativa, ya que se desconoce si el registro se puede comprimir hasta que se comprueba el cuarto bloque. Por esta razón, se propone el uso de un Buffer de Registro Destino (BRD), tan grande como una entrada del slice, el cual almacena los cuatro bloques de la unidad SIMD mientras se determina la compresibilidad del registro actual. Cuando la compresión no es posible, se notifica a la USR cambiando el estado del bit $c_{compr}$ y la segmentación se detiene durante cuatro ciclos hasta que el registro completo del BRD se escribe en la entrada del slice correspondiente.




\subsection{Unidad de Selección de Redirección (USR)}

La USR consta de un mapa de bits de tamaño $256\times 4$ refiriéndose a cada bloque del slice y dos codificadores de prioridad con 1024 y 256 entradas cada uno para seleccionar entradas defectuosas y fiables, respectivamente. El mapa de bits se preconfigura con el mapa de fallos del escenario de fiabilidad elegido y se actualiza a lo largo de la ejecución de una aplicación con las redirecciones ocupadas y liberadas. Por supuesto, los bloques defectuosos se marcan permanentemente como ocupados en el mapa de bits y no se  pueden utilizar. Según el estado del mapa de bits, el codificador de 1024 entradas selecciona un bloque libre de una entrada defectuosa para redireccionar un registro comprimido, mientras que el codificador de 256 entradas selecciona una entrada de registro fiable, es decir cuatro bloques libres, para redireccionar un registro incompresible. La USR escoge el tipo de redirección en función del estado de compresión del registro destino (bits $c_{compr}$). El resto de las entradas de la USR son el contenido TR del registro destino.


\begin{table*}[t!]
\renewcommand{\baselinestretch}{1}
\small
\centering
\caption{Temporización, energía, potencia y área para un nodo de fabricación de 28 nm y una frecuencia de reloj de 1 GHz. N/A: No aplicable.}
\begin{tabular}{|c|c|c|c|c||c|c||c|c|} \cline{2-9}
\multicolumn{1}{c|}{} & Slice  & Slice  & Slice  & Slice  & Com & Des & TR & USR\\
\multicolumn{1}{c|}{} & @Conv & @Com & @Agr & @Dis &   &     &     &  \\\hline\hline
Tiempo de acceso (ns) & \multicolumn{4}{c||}{0,85} & 0,95 & 0,95 & 0,09 & 0,11 \\\hline
Energía por lectura (pJ)    & 247,38 & 84,38 & 84,90 & 68,25 & N/A & 0,62 & 0,54 &  N/A\\\hline
Energía por escritura (pJ)    & 302,23 & 97,68 & 99,76 & 78,33 & 0,72 & N/A & 0,50 &  0,22\\\hline
Potencia estática (mW)    & 58,58 & 30,79 & 35,18 & 27,73 & 6,67 & 7,14 & 0,41 &  15,41\\\hline
Área ($mm^{2}$) & \multicolumn{4}{c||}{0,680} & 0,005 & 0,007 & 0,005 & 0,014\\\hline
\end{tabular}
\label{tvalues}
\end{table*}

Para cada instrucción con un registro destino, la USR permite que uno de los codificadores obtenga de manera preventiva una redirección mientras la instrucción atraviesa el pipeline. Esta redirección preventiva se refiere a una redirección complementaria con respecto al estado del registro actual. Por ejemplo, si la redirección actual se refiere a una entrada fiable ($c_{tr}=0$), la USR selecciona preventivamente un bloque fiable de una entrada defectuosa, asumiendo que el estado de compresión cambiará. De este modo, cuando el primer bloque del registro alcanza la etapa WB, la USR compara los bits $c_{tr}$ y $c_{compr}$, y acciona el puerto de escritura en consecuencia (bit $rsel$), ya sea seleccionando la nueva redirección ($entrada_{usr \; {dest}}$ y $blq_{usr \; {dest}}$) o la redirección anterior de la TR.

Las redirecciones preventivas permiten situar a la USR en la etapa WB sin aumentar la profundidad del pipeline. Si se escribe un registro por primera vez ($v_{dest}=0$), la USR permite a ambos codificadores obtener redirecciones preventivas. Una vez que la instrucción abandona el pipeline, la USR libera la redirección no utilizada y actualiza la TR si es necesario.


En el caso de un fallo de compresión, la USR selecciona la redirección correspondiente a una entrada fiable y los bloques no comprimidos del BRD se escriben en el slice uno tras otro ($misp=1$).
También, cuando una instrucción escribe en un registro de respaldo ($m_{dest}=1$), la USR trata de encontrar una nueva redirección en el slice para el registro con el fin de mitigar las penalizaciones de latencia por el acceso a la LDS, obteniendo dos redirecciones preventivas al igual que en la primera escritura a un registro. Por otro lado, cuando no se puede encontrar una nueva redirección en el slice, la USR establece el bit $m_{dest}$ a `1' y actualiza la $entrada_{tr \; {dest}}$ con el desplazamiento de la partición de la LDS donde se asigna el registro.


Finalmente, cuando un \emph{wavefront} libera su ventana de registros porque ha terminado la ejecución, la USR accede a la ventana del \emph{wavefront} en la TR, invalidando todas las filas ($v=0$) del \emph{wavefront} y liberando todas las redirecciones correspondientes en el mapa de bits de la USR.

\subsection{Retardo, Energía, Potencia y Área}
\label{over}

Esta sección estima el retardo, energía, potencia y área de los componentes principales de RRCD. Las estructuras de memoria como el slice y la TR se han modelado con CACTI 7.0~\cite{cactip}, mientras que la lógica combinacional y los flip-flops presentes en las unidades Com/Des y la USR se han sintetizado con Synopsis Design Compiler y simulado con Mentor Graphics Modelsim. La librería corresponde a una tecnología de 28 nm de bajo $V_{th}$ disponible en el ámbito académico. Se asume la frecuencia de reloj de 1 GHz de la GPU AMD GCN HD 7770. Se trata de la GPU considerada en todos los experimentos.


La Tabla~\ref{tvalues} muestra los resultados. El tiempo de acceso y la energía dinámica (lecturas y escrituras) del slice se refiere al acceso a un bloque de 64 bytes. Para las unidades Com/Des, TR y USR, estos parámetros se refieren a la evaluación de un bloque, el acceso a una entrada y la obtención de una nueva redirección, respectivamente. Sin embargo, en el caso de la USR, el tiempo de acceso se refiere exclusivamente al envío de la nueva redirección al puerto de escritura en la propia etapa del pipeline, ya que las redirecciones preventivas se obtienen en etapas anteriores. Nótese que el consumo dinámico no es aplicable en algunos componentes de RRCD debido a que no intervienen en un acceso de lectura o escritura. Los resultados del slice se muestran para los tres escenarios de fiabilidad estudiados más un escenario convencional donde la GPU trabaja con una $V_{dd}$ nominal para evitar fallos. Al igual que la GPU convencional, todos los componentes de RRCD se mantienen a $V_{dd}$ nominal por la misma razón.



Se asume que el tiempo de acceso del slice permanece constante independientemente del valor de $V_{dd}$. Sin embargo, la reducción de $V_{dd}$ puede conducir a un aumento del retardo de conmutación de los transistores~\cite{energy2}. En este sentido, se ha medido la pérdida media de rendimiento al segmentar el acceso al slice y aumentar la latencia de 1 a 3 ciclos adicionales. De forma similar a~\cite{wctc,sttcomp}, el impacto en el rendimiento oscila entre un 0,6 y un 1,5\%.
Estos resultados indican que la ampliación de la profundidad del pipeline tiene un impacto relativamente bajo en el rendimiento, en particular si hay suficiente paralelismo a nivel de hilos de los \emph{wavefronts} asignados al slice.


La energía dinámica y la potencia estática del slice disminuyen con $V_{dd}$. Las mayores reducciones se observan en la energía dinámica, puesto que aumenta cuadráticamente con $V_{dd}$.
En comparación con el slice, los componentes de RRCD reducen en gran medida la energía, potencia y área. La potencia es el parámetro que más afecta, nótese que la potencia de todos los componentes de RRCD se sitúa en 36,77 mW, lo que supone casi dos tercios de la potencia del slice cuando opera con una $V_{dd}$ nominal. Por otro lado, la estimación del área conjunta de todos los componentes propuestos es 0,038 $mm^2$, lo que corresponde a un 5,6\% del slice.


Recuérdese que el tiempo de acceso de las unidades Com/Des (0,95 ns) obliga a ampliar el pipeline en dos etapas adicionales. Por el contrario, el tiempo de acceso es lo suficientemente pequeño tanto en la TR (0,09 ns) como en la USR (0,11 ns) como para encajarlas en las etapas originales de traducción y \emph{writeback}. Además, el impacto de los muxes 2:1 adicionales es mínimo, puesto que el retardo de este componente es 13 ps de acuerdo con Synopsis DC Ultra.


\section{Evaluación Experimental}
\label{exp}

\begin{table}[t!]
\renewcommand{\baselinestretch}{1}
\small
\centering
\caption{Configuración de la GPU y jerarquía de memoria.\label{param}}
{
\begin{tabular}{|l|l|} \hline
Frecuencia de reloj        &   1 GHz \\
CUs                        & 10 \\
Slice                      & 64 KB, 4/CU, 4-1-1-1 ciclos/instr. \\
Máx. WFs/CU     & 16 \\
Hilos/WF   & 64\\
\hline\hline
Todas las cache                 &   LRU, 64B/línea \\
Caches L1 escalares        &   16 KB, 4-vías, 1/CU, 1 ciclo \\
Caches L1 texturas          &   16 KB, 4-vías, 1/CU, 1 ciclo \\
Caches LDS                 &   64 KB \emph{scratch}, 1/CU, 1 ciclo \\
2$\times$ caches L2                  &   128 KB, 16-vías/módulo, 10 ciclos \\
Memoria principal               &   2 canales/módulo L2, 100 ciclos \\
\hline
\end{tabular}}
\end{table}

RRCD se ha modelado con el simulador ciclo-a-ciclo Multi2Sim~\cite{m2sgpu}. Los resultados incluyen el tiempo de ejecución de las aplicaciones y estadísticas adicionales del procesador requeridas para estimar el consumo de energía. La energía total se ha calculado combinando estadísticas de Multi2Sim con números de energía provenientes de CACTI y Synopsis. La Tabla~\ref{param} muestra los parámetros de configuración y jerarquía de memoria de la GPU modelada. Todos los benchmarks se ejecutan de principio a fin.

\subsection{Impacto en el Rendimiento}
\label{expperf}

Para entender mejor las fuentes principales de degradación de rendimiento, en primer lugar se clasifican las escrituras en el slice como accesos regulares, escrituras que requieren una nueva redirección y escrituras que disparan un acceso hacia la LDS debido a registros de respaldo. A continuación, se analiza el impacto en el rendimiento del sistema.

\subsubsection{\emph{Desglose de Escrituras en Slice}}

\begin{figure}[t!]
\centering
\includegraphics[width=0.99\columnwidth]{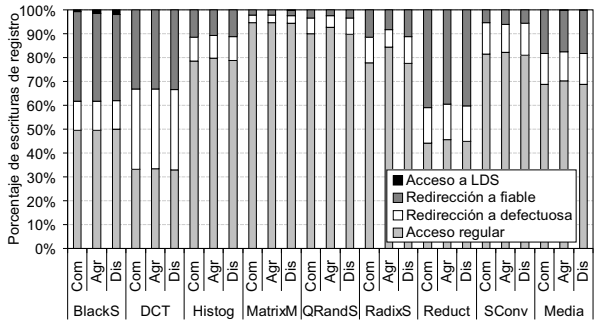}
\caption{Porcentaje de escrituras de registro que requieren un acceso regular sin redirección, una redirección hacia una entrada fiable o defectuosa y accesos adicionales a la LDS.}
\label{exppatch}
\end{figure}

La Figura~\ref{exppatch} ilustra un desglose de tipos de escritura en el slice. Nótese que la primera escritura de un registro siempre se considera como una redirección hacia una entrada fiable o defectuosa, o un acceso a la LDS.

Los resultados muestran que, en aplicaciones como \emph{QRandS} y \emph{MatrixM}, el porcentaje de escrituras normales sin redirección alcanza el 90\%. Este porcentaje es en promedio del 70\%, lo que significa que la mayoría de los registros no cambian el estado de compresión durante toda su vida útil. En otras palabras, una vez que RRCD selecciona una entrada donde redireccionar un registro, normalmente se refiere al mismo registro lógico durante toda la ejecución de un \emph{wavefront}. El hecho de que los resultados sean bastante homogéneos para una aplicación dada bajo los diferentes escenarios de fiabilidad también confirma este razonamiento.

RRCD proporciona suficientes entradas para  redireccionar registros en todos los escenarios de fiabilidad estudiados, ya que las escrituras hacia la LDS sólo son notables en \emph{BlackS}, donde el porcentaje oscila entre el 1-2\%. La combinación de dos factores clave son la causa de la presencia de escrituras en la LDS para esta aplicación. Primero, el porcentaje de registros comprimidos es bastante bajo, lo que implica que una gran cantidad de entradas defectuosas no se pueden explotar con fines de redirección. En segundo lugar, la utilización del banco de registros es muy alta (véase la Figura~\ref{motfig}), lo cual limita las oportunidades de encontrar una entrada redireccionable.

\subsubsection{\emph{Impacto en el Rendimiento del Sistema}}
\label{systperf}

\begin{figure}[t!]
\centering
\includegraphics[width=0.99\columnwidth]{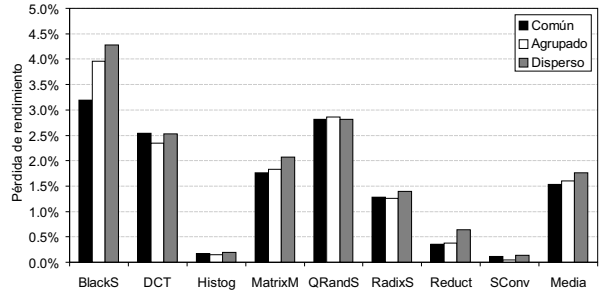}
\caption{Pérdida de rendimiento de RRCD frente a un banco de registros convencional.
}
\label{slowd}
\end{figure}

La Figura~\ref{slowd} ilustra la degradación de rendimiento del sistema para la técnica RRCD con respecto a un banco de registros convencional operando en la tensión nominal. Nótese que, aparte de los accesos adicionales a la LDS, el aumento de la profundidad del pipeline, así como las especulaciones erróneas en las operaciones de compresión también pueden afectar al rendimiento. En este sentido, la predicción del mecanismo de compresión es muy precisa, puesto que el porcentaje de escrituras con una especulación incorrecta es un 2.4\% en promedio.

El impacto en el rendimiento varía ampliamente entre las aplicaciones. \emph{BlackS} muestra el mayor impacto en el rendimiento, puesto que este benchmark sufre los tres efectos mencionados anteriormente. Sin embargo, la pérdida de rendimiento no excede un 4.5\% con respecto al diseño convencional. En 
\emph{DCT}, \emph{Histog} y \emph{QRandS}, el impacto en el rendimiento es bastante uniforme con independencia de los escenarios de fiabilidad. Estas pérdidas se atribuyen al efecto combinado de la especulación errónea y el aumento de etapas en el pipeline. El resto de aplicaciones se encuentran afectadas principalmente por la mayor longitud del pipeline. Sin embargo, los resultados confirman que el planificador es capaz de despachar instrucciones independientes de \emph{wavefronts} diferentes de manera frecuente, enmascarando los efectos de un pipeline más profundo en el rendimiento.

\begin{figure*}[!t]
\centering
\includegraphics[width=0.9\textwidth]{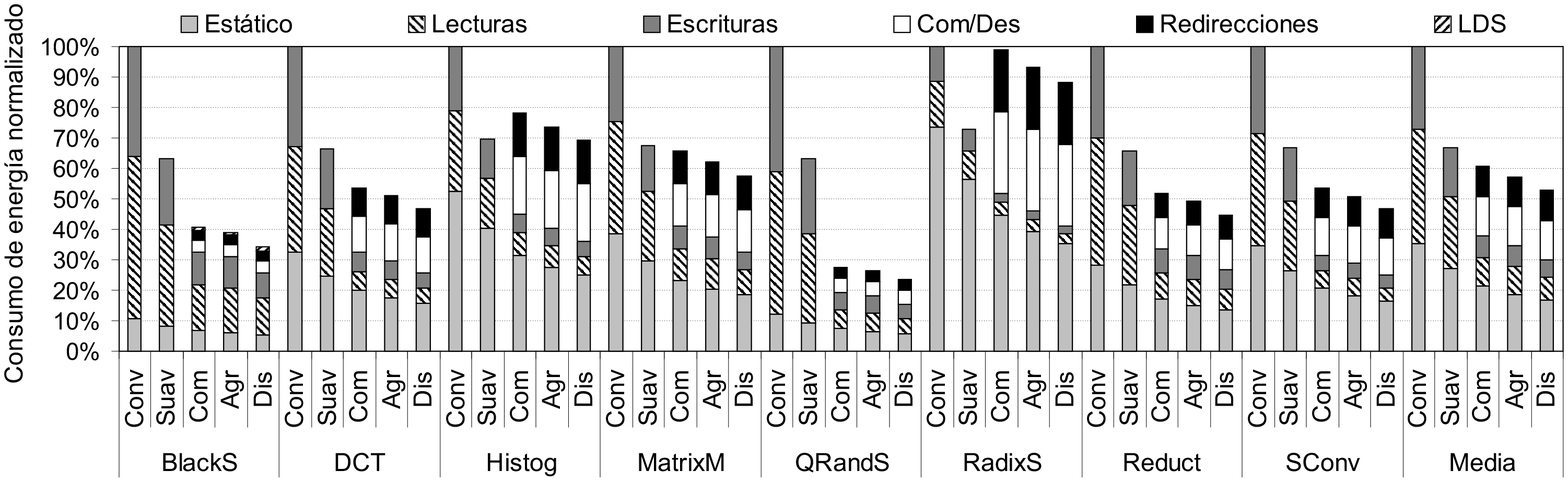}
\caption{Consumo de energía normalizado para RRCD y un esquema de suavizado de ruido de tensión (Suav) con respecto a un banco de registros convencional.}
\label{energy}
\end{figure*}

En resumen, el impacto en el rendimiento del mecanismo RRCD es en promedio 1,53, 1,60 y 1,76\% para \emph{Común}, \emph{Agrupado} y \emph{Disperso}, respectivamente, frente a un banco de registros convencional.

\subsection{Ahorro Energético}
\label{expenergy}

La Figura~\ref{energy} muestra el consumo de energía normalizado del slice con respecto al diseño convencional. Con fines comparativos, se incluye también un esquema de suavizado de ruido de tensión (Suav) operando con una $V_{dd}=600$ mV. La energía se divide en coste estático y dinámico. A su vez, la energía dinámica se clasifica en coste debido a operaciones de lectura y escritura en el slice. La sobrecarga de energía de las especulaciones erróneas se agrega a la categoría de escritura. La etiqueta \emph{Com/Des} se refiere al consumo estático y dinámico de las unidades de (des)compresión, mientras que el consumo estático y dinámico de la TR, USR y BRD se acumula en la categoría de \emph{Redirecciones}. Finalmente, también se cuantifica la sobrecarga de energía dinámica por acceso a la LDS debido a registros de respaldo.

La contribución del coste estático sobre la energía total varía ampliamente entre aplicaciones. Por ejemplo, las aplicaciones con un uso del banco de registros relativamente bajo muestran una contribución menor de consumo dinámico, pero siguen acumulando energía estática en cada ciclo. Este es el caso de \emph{RadixS}, donde se asigna una gran cantidad de \emph{wavefronts} durante la ejecución de la aplicación, pero el número de accesos al slice es relativamente bajo. Dado que el consumo estático aumenta linealmente con $V_{dd}$, Suav y RRCD reducen significativamente este consumo en comparación con el esquema convencional. RRCD logra tal ahorro de energía a pesar de la sobrecarga de energía estática debido a un mayor tiempo de ejecución frente Conv y Suav. Como se esperaba, bajo un escenario \emph{Disperso}, con la $V_{dd}$ más baja, se obtiene el mayor ahorro de energía estática, seguido de \emph{Común} y \emph{Agrupado}.

La reducción de $V_{dd}$ tiene un mayor impacto en el coste dinámico, ya que este parámetro tiene un efecto cuadrático sobre este tipo de consumo. Además, RRCD también contribuye a reducir aún más el consumo dinámico frente a Conv y Suav. Esto se debe a que, en un acceso de lectura/escritura a un registro comprimido del slice, sólo se accede a un bloque que contiene los datos comprimidos en lugar de al registro completo. Las aplicaciones con un gran uso del banco de registros y una capacidad elevada de compresión, como \emph{QRandS} y \emph{SConv}, muestran un gran ahorro de energía dinámica.

Recuérdese que los componentes de RRCD operan con una $V_{dd}$ nominal para evitar fallos. Esto impone una sobrecarga de energía, pero no impide que RRCD reduzca en gran medida el consumo total de energía gracias a: i) una reducción agresiva de $V_{dd}$ en el slice y ii) una mitigación del coste dinámico del slice al acceder a registros comprimidos. En comparación con Suav, sólo \emph{Histog} y \emph{RadixS} muestran un mayor consumo. Esto se debe principalmente a que estas aplicaciones hacen un uso del slice relativamente bajo y a las pocas oportunidades de compresión de datos.

Nótese que el coste de acceso a la partición de respaldo de la LDS sólo se aprecia levemente en \emph{BlackS}, confirmando que el número de accesos a registros de respaldo es mucho menor que el número de accesos a registros redireccionados en el propio slice.

En resumen, el ahorro total de energía de RRCD bajo los escenarios \emph{Común}, \emph{Agrupado} y \emph{Disperso} es de 39, 43 y 47\%, respectivamente, con respecto al esquema convencional. Comparado con Suav, estos porcentajes son 10, 14 y 21\%.

\section{Trabajo Relacionado}
\label{related}

Esta sección describe trabajos recientes que utilizan la compresión de datos en el banco de registros GPU, así como técnicas de redirección de registros para tolerar fallos permanentes en esta estructura de memoria.

\subsection{Compresión de Datos en Bancos de Registros}

Zhang \emph{et al}. implementan un banco de registros con tecnología \emph{spin-transfer torque magnetic} RAM~\cite{sttcomp}. Con el objetivo de reducir la energía dinámica y las latencias prolongadas de las operaciones de escritura, los autores utilizan el algoritmo BDI para comprimir registros.

Warped-Compression ahorra energía estática en el banco de registros al aprovechar la compresión BDI~\cite{wctc}. Para aquellos registros identificados como compresibles, este enfoque mantiene los datos comprimidos en los bits menos significativos de estos registros, mientras que las celdas con los bits restantes que no contienen datos útiles se apagan mediante la técnica \emph{power gating}.

La propuesta EREER elimina las componentes adyacentes y duplicadas de un registro, conservando las componentes no duplicadas y los bits de control necesarios para descomprimir el registro en los bits menos significativos de la entrada~\cite{ereer}. Las celdas que almacenan las componentes no utilizadas se apagan para reducir el consumo de energía.

\emph{Power gating} ayuda no sólo a reducir el consumo estático, sino también a mitigar el efecto de envejecimiento \emph{Negative Bias Temperature Instability} (NBTI). A diferencia de Warped-Compression y EREER, RC+RAR es una técnica que desconecta registros completos almacenando los datos comprimidos en una memoria pequeña y auxiliar libre del efecto  NBTI por diseño~\cite{valerotc}.

\subsection{Técnicas de Redirección}

GR-Guard es una técnica de redirección que aprovecha las entradas fiables y que contienen registros inútiles para almacenar registros útiles de entradas defectuosas, evitando el uso de dichas entradas defectuosas~\cite{Tan2016}. Una entrada de registro almacena datos inútiles durante el periodo de tiempo desde la última lectura hasta la siguiente operación de escritura. Dado que esta información es desconocida en tiempo de ejecución, GR-Guard hace uso del compilador y modifica el repertorio de instrucciones para identificar estas entradas de registro en tiempo de ejecución. Específicamente, el formato de instrucción incluye un bit adicional para cada operando,  distinguiendo de esta manera si la entrada de registro asociada es útil o no lo es. Al contrario que GR-Guard, el presente trabajo no explota el compilador ni modifica el repertorio de instrucciones. RRCD se basa en la observación de que los contenidos del registro se pueden comprimir fácilmente en tiempo de ejecución, lo que permite tratar las entradas defectuosas como ubicaciones para la redirección en sí mismas, aumentando las oportunidades de redirección con respecto a GR-Guard.

iPatch tolera fallos en las cache L1 de la CPU explotando la replicación inherente de los contenidos de la cache en otras estructuras del pipeline, como la \emph{trace} cache, el MSHR y la cola de escrituras~\cite{Ipatch}. La propuesta almacena las líneas de cache L1 defectuosas en dichas estructuras del pipeline. Esta técnica requiere modificaciones en los componentes del procesador, incluida la gestión de la consistencia de memoria en las estructuras utilizadas como respaldo, lo cual complica el diseño y la verificación del procesador. Además, la cobertura de fallos se limita a las cache L1 debido al tamaño relativamente pequeño de las estructuras de respaldo.

\section{Conclusiones}
\label{conclus}

El objetivo principal de este trabajo ha sido asegurar la tolerancia a fallos permanentes de un banco de registros GPU operando por debajo del límite seguro de tensión de alimentación. Para ello, este trabajo ha propuesto una técnica de redirección, RRCD, que aprovecha la redundancia inherente de los datos de las aplicaciones GPGPU para comprimir entradas de registro en tiempo de ejecución. Con la compresión habilitada, las entradas de registro defectuosas vuelven a ser útiles porque pueden almacenar múltiples registros comprimidos en el resto de sus celdas fiables, mejorando de manera efectiva la capacidad de almacenamiento del banco de registros.

Los resultados experimentales han mostrado que RRCD garantiza un funcionamiento fiable de un banco de registros con un 39\% de entradas defectuosas. Para esta tasa de fallos, el consumo de energía se reduce en un 47 y un 21\% en comparación con un banco de registros libre de fallos que opera con una tensión nominal y en el límite seguro de tensión de alimentación, respectivamente, mientras que el impacto en el rendimiento y el área permanece por debajo del 2 y 6\%.

\section*{Agradecimientos}

Este trabajo ha sido financiado por la Universidad de Zaragoza bajo el proyecto JIUZ-2019-TEC-08, MINECO/AEI/FEDER (UE) bajo los proyectos TIN2016-76635-C2-1-R y PID2019-105660RB-C21, Gobierno de Aragón (Grupo T58\_20R) y FEDER 2014-2020 \emph{Construyendo Europa desde Aragón}.

\bibliographystyle{abbrv}
\bibliography{./main}

\end{document}